\documentclass{article}
\usepackage{spconf,amssymb, amsmath, graphicx, paralist,subfigure,epsfig,cite}
\usepackage{float}
\usepackage[ruled,vlined]{algorithm2e}
\usepackage[usenames,dvipsnames]{color}
\usepackage{caption}
\newtheorem{theorem}{Theorem}

\newtheorem{cor}{Corollary}

\def\mb{\mathbf}
\def\mbb{\mathbb}
\def\mc{\mathcal}

\title{Analysis of Contractions in System Graphs:\\
	Application to State Estimation}
\name{%
\begin{tabular}{@{}c@{}}
	Mohammadreza~Doostmohammadian$^{\dagger \star}$, Themistoklis Charalambous$^\dagger$,~\textit{Senior Member,~IEEE,} \qquad \\
	Miadreza Shafie-khah$^\ast$, Hamid R. Rabiee$^\diamond$,\textit{Senior Member,~IEEE,} \qquad \\ and~Usman~A.~Khan$^\ddagger$,~\textit{Senior Member,~IEEE }
\end{tabular}
\thanks{This work is partially supported by National Science Foundation under awards~\#1903972 and~\#1935555. Corresponding author email: {\tt\small doost@semnan.ac.ir, mohammadreza.doostmohammadian@aalto.fi}}}

\address{
	$\dagger$ School of Electrical Engineering, Aalto University, Espoo, Finland.
	\\ 
	$^\star$ Faculty of Mechanical Engineering, Semnan University, Semnan, Iran. \\
	$^\ast$ School of Technology and Innovations, University of Vaasa, Vaasa, Finland
	\\
	$^\diamond$ Computer Engineering Department, Sharif University of Technology, Tehran, Iran \\
	$^\ddagger$Electrical and Computer Engineering Department, Tufts University, Medford, MA, USA.}

\begin{document}
\maketitle

%
%
%
%
\begin{abstract}
Observability and estimation are closely tied to the system structure, which can be visualized as a \textit{system graph}--a  graph that captures the inter-dependencies within the state variables. For example, in social system graphs such inter-dependencies represent the social interactions of different individuals. It was recently shown that contractions, a key concept from graph theory, in the system graph are critical to system observability, as (at least) one state measurement in every contraction is necessary for observability. Thus, the size and number of contractions are critical in recovering for loss of observability. In this paper, the correlation between the average-size/number of contractions and the global clustering coefficient (GCC) of the system graph is studied. Our empirical results show that estimating systems with high GCC requires fewer measurements, and in case of measurement failure, there are fewer possible options to find substitute measurement that recovers the system's observability. This is significant as by tuning the GCC, we can improve the observability properties of large-scale engineered networks, such as social networks and smart grid. 
\end{abstract}


%
%
%
%
\section{Introduction} 
\label{sec_intro}
Large-scale networked systems have seen a surge of interest in recent control and signal processing literature with applications in IoT and CPS ~\cite{csl2020,rana2017distributed,arasteh2016iot}. A key challenge in such networks is state estimation~\cite{SPL17,rana2017distributed,kabiri2018enhancing} via a distributed network of measurements. From this perspective of distributed estimation, an effective tool is the system graph in which nodes represent state variables and edges between two nodes show coupling among the two state variables~\cite{woude:03,Liu_nature,ortega2018graph}, motivating structural control and graph signal processing. In this sense, structural observability is related to certain system graph properties relying only on the system structure, and not on the exact system parameter values~\cite{woude:03,SPL17,Liu_nature,carvalho2017composability}.

An important graph-theoretic property to study system observability is the notion of \textit{contraction} in the system graph, which is the dual of \textit{dilation} in controllability~\cite{Liu_nature}. In a contraction, multiple nodes are contracted (connected) to a fewer group of nodes. It is known that measuring one state node in every contraction is essential for network observability~\cite{TNSE17,commault2011sensor}. All states in a contraction are thus observationally equivalent which is significant for observability recovery, for example, in  sensor/measurement failure~\cite{SPL17,commault2011sensor}. The size of contractions is also a key property, representing the number of possible options for estimation/observability recovery. A large contraction presents more choices of equivalent state measurements to replace the failed/faulty observation, or, for example, to minimize cost~\cite{guo2020actuator,moothedath2018minimum,spl18}. Also, the number of contractions represents the number of necessary measurement (or sensors)  for estimation. The size and distribution of contractions in a system graph depend on certain graph properties. This work particularly studies how {the} \textit{global clustering coefficient} affects the distribution of contractions. This paper is a nonlinear model extension of our previous works~\cite{TNSE17,csl2020} on \textit{local clustering coefficient} and \textit{degree heterogeneity}.

This paper models the nonlinear system as a random Scale-Free (SF) graph. The reason is that the structure of most real-world systems resemble the structure of SF graphs~\cite{newman2006structure}. To study the effect of {the} GCC, as our main contribution, the distribution of  size/number of contractions in SF graphs and clustered SF (CSF) graphs are compared. Due to specific formation in CSF graphs (known as \textit{triad formation}) they have higher GCC, while their other properties (particularly power-law degree distribution) are similar to SF graphs. The significance of this contribution is that by tuning the network GCC, e.g., adopting the results of~\cite{islam2018towards,moore2021inclusivity}, one can improve/impair system observability properties.  Our results can be used in {the} design of large-scale man-made networks to improve their estimation properties in terms of reducing necessary observer nodes (sensor locations) for cost-optimal estimation. An example of such re-design of power grid is given in Section~\ref{sec_power} as another contribution of this paper. Another possible application is in changing the structure of social networks  to hinder the possibility of distributed estimation~\cite{montijano2020distributed,isj2020} and, therefore, improve information privacy and reduce the vulnerability towards information leakage~\cite{lovato2020distributed}. 

The rest of this paper is as follows. Section~\ref{sec_cont} describes graph-theory notions to define the contractions. Section~\ref{sec_sys} states the specific application to system estimation and observability. In Section~\ref{sec_SF}, the distribution of contractions in SF and CSF graphs are compared, and an illustrative example application in power grid monitoring is given in Section~\ref{sec_power}. Conclusions and future research are presented in Section~\ref{sec_conc}.

\vspace{-0.25cm}
\section{Contractions in Graphs} \label{sec_cont}
\vspace{-0.25cm}
We consider the complex system, for example a social system, as an undirected graph or a strongly-connected directed graph (SC digraph) denoted by~$\mc{G}=(\mc{V},\mc{E})$, with the node set $\mc{V}$ (representing the $n$ states) and edge/link set $\mc{E}=\{(v_i,v_j)\}$. The associated bipartite system graph~$\Gamma=(\mc{V}^+,\mc{V}^-,\mc{E}_\Gamma)$ is defined with two disjoint left/right node sets~$\mc{V}^+$ and~$\mc{V}^-$, and edges~$\mc{E}_\Gamma=\{(v_j^-,v_i^+)|(v_j,v_i) \in \mc{E}\}$. In~$\mc{G}$, the edges with no common end node are called a matching~$\underline{\mc{M}}$, which equivalently  in~$\Gamma$ represent the subset of edges not incident on the same node in $\mc{V}^+$. In other words, $\underline{\mc{M}}$ is a set of pairwise disjoint edges (with no loop).   A matching with maximum size is called maximum (cardinality) matching~$\mc{M}$, which is not a subset of any other matching. Note that there are many possible choices of $\mc{M}$ in general. The nodes respectively in~$\mc{V}^+$ and~$\mc{V}^-$ incident to the chosen~$\mc{M}$ are  denoted by $\partial \mc{M}^+$ and~$\partial \mc{M}^-$, and the nodes in ${\delta \mc{M} = \mc{V}^+ \backslash \partial \mc{M}^+}$ are unmatched  in~$\mc{V}^+$. Given~$\mc{M}$, let~$\Gamma^\mc{M}$ be the auxiliary graph made by reversing all edges in~$\mc{M}$, and holding all the other edges~$\mc{E}_{\Gamma} \backslash \mc{M}$ in~$\Gamma$. In $\Gamma^\mc{M}$, an alternating path associated to~${\mc{M}}$ (also called ${\mc{M}}$-alternating path), denoted by$\mc{Q}_{\mc{M}}$, is a path starting from a node in~$\delta \mc{M}$ with its edges alternately in $\mc{M}$ and not in $\mc{M}$.
An augmenting path $\mc{P}_{\mc{M}}$ associated to~${\mc{M}}$ (also called ${\mc{M}}$-augmenting path) is an ${\mc{M}}$-alternating path in $\Gamma^\mc{M}$ that starts from and ends in $\delta \mc{M}$. For a matching $\underline{\mc{M}}$ and associated $\mc{P}_{\underline{\mc{M}}}$, $\underline{\mc{M}} \oplus \mc{P}_{\underline{\mc{M}}}$ represents a new matching with one more edge than $\underline{\mc{M}}$, where $\oplus$ is the XOR operator.
\textit{In~$\Gamma^\mc{M}$, a contraction $\mc{C}_j$, associated to an unmatched node~$v_j \in \delta \mc{M}$, is defined as the set of all state nodes in $\mc{V}^+$ reachable by ${\mc{M}}$-alternating paths starting from~$v_j$}. Intuitively speaking, contraction  represents subset of nodes  linking to smaller subset of nodes \cite{murota,TNSE17}.
Algorithm~1 {\cite{murota,TNSE17}} presents the pseudo-code {for finding} graph contractions with polynomial order complexity $\mc{O}(n^{2.5})$. {Polynomial complexity facilitates} applications in large-scale as in social networks or power grids. 
\vspace{-0.2cm}
\begin{algorithm} \label{alg_cont}
	\textbf{Given:} System graph $\mc{G}$ 
	
	Find $\Gamma$\;
	Find $\underline{\mc{M}}$ \;
	Find $\Gamma^{\underline{\mc{M}}}$ \;
	\While{ $\mc{P}_{\underline{\mc{M}}}$ exist}{
		\For{nodes in $\delta \underline{\mc{M}}$}{  
			Find $\mc{P}_{\underline{\mc{M}}}$ \;
			$\underline{\mc{M}} = \underline{\mc{M}} \oplus \mc{P}_{\underline{\mc{M}}}$ \;
		}
	}
	Find  $\Gamma^{\mc{M}}$ \;
	\For{state nodes in $\delta \mc{M}$}{  
		Find $\mc{Q}_{\mc{M}}$ in $\Gamma^{\mc{M}}$ \;
		Put nodes in $\mc{V}^+$ reachable by $\mc{Q}_{\mc{M}}$ in $\mc{C}_i$\;}
	
	\textbf{Return} $\mc{C}_i, i = \{1,...,l\}$\;\
	
	\caption{\small{Finding contractions in a graph \cite{TNSE17,murota}.}}
\end{algorithm}
\vspace{-0.2cm}

In this paper, as our main contribution, we aim to understand possible correlation between the GCC and prevalence of contractions, and  interpret the implication of this relation {through a} system estimation perspective.

\section{Application to state estimation} \label{sec_sys}
In this work, a \textit{nonlinear} autonomous dynamic system (in contrast to the linear model in \cite{TNSE17}) is considered as,
\begin{eqnarray}\label{eq_sys_nonlin}
\dot{\mb{x}} &=& f(\mb{x}(t))+ \mb{v},
\end{eqnarray}
where the state variable $\mb{x}=[x^{1},\ldots,x^{n}]^\top\in\mbb{R}^n$ is to be estimated via the measurements,
\begin{eqnarray}\label{eq_sys_nonlin2}
\mb{y}(t) &=& g(\mb{x}(t)) + \mb{r},
\end{eqnarray}
where $\mb{y}=[y^1,\ldots,y^m]\in\mbb{R}^m$ is the measurement, and~$\mb{v}$ and~$\mb{r}$ are Gaussian noise. The system model~\eqref{eq_sys_nonlin}-\eqref{eq_sys_nonlin2} can be represented as a Linear-Structure-Invariant (LSI) model as,
\begin{eqnarray}\label{eq_sys1}
\dot{\mb{x}} &=& A(t)\mb{x}(t) + \mb{v},
\\ \label{eq_sys2}
\mb{y}(t) &=& C(t)\mb{x}(t) + \mb{r},
\end{eqnarray}
where $A(t)$ and $C(t)$ are time-dependent system and measurement matrices representing the linearization of the system and measurement functions $f(\cdot)$  and $g(\cdot)$ over time. Recall that, from Kalman filtering theory, the underlying system can be estimated if it is \textit{observable} via the given measurements. System observability implies that the global vector,~$\mb{x}$, can be uniquely determined by the measurements,~$\mb{y}$. As shown in~\cite{Liu_nature}, the observability of the nonlinear model \eqref{eq_sys_nonlin}-\eqref{eq_sys_nonlin2} is equivalent with the observability of the linearized model \eqref{eq_sys1}-\eqref{eq_sys2} over all operating points. The structure (the zero-nonzero pattern) of the associated linearized matrices $A(t)$ and $C(t)$ are time-invariant while the numerical values of their nonzero entries may vary at different operating points, implying the name structure-invariant. This motivates the concept of \textit{structural observability} (or \textit{generic observability}) based on structured systems theory~\cite{woude:03,Liu_nature,carvalho2017composability}, which provides a graph-theoretic method to check for system observability. In structural analysis the system is modeled as a \emph{system graph}, where a node $v_i$ models a state $x^i$ and a link $v_j \rightarrow v_i$ models the dependency of the two state variables $x^i$ and  $x^j$. In other words, if $f^i$ is a function of $x^j$ then the entry $\frac{\partial f^i}{\partial x^j}$ in linearized matrix $A$ is nonzero while its exact value depends on the operating point and may change over time~\cite{isj2020,carvalho2017composability}. Denote the system graph by $\mc{G}_A=(\mc{V},\mc{E}_A)$, with state nodes $\mc{V}$ and  $\mc{E}_A=\{(v_j,v_i)~|~\frac{\partial f^i}{\partial x^j}\neq0\}$ including the  edges  $v_j\rightarrow v_i$.
Using the definitions in Section~\ref{sec_cont}, the next theorem states necessary conditions for  observability of the system graph $\mc{G}_A$.

\begin{theorem} \label{thm_unmatched}
	Let $\delta \mc{M}$ denote the set of unmatched nodes of system graph $\mc{G}_A$ associated with an autonomous LSI system. To ensure observability, it is necessary to measure every unmatched state in $\delta \mc{M}$.
\end{theorem}
We refer to~\cite{Liu_nature} for the proof (in the dual case of controllability). Based on the definition, for a given maximum matching $\mc{M}$, every node $v_j \in \delta \mc{M}$ belongs to a contraction $\mc{C}_i$, while the nodes $\mc{C}_i \backslash v_j$ are all matched. This leads to the following \textit{observational equivalence} property in contractions:

\begin{theorem} \label{thm_contr}
	Consider an LSI system abstracted as a graph (undirected or SC) with   contractions $\mc{C}=\{\mc{C}_1,...,\mc{C}_m\}$. The necessary  condition for observability is to measure (at least) one state node in every contraction. 
\end{theorem}
\begin{cor} \label{cor_equiv}
    Nodes in a contraction are \textit{equivalent} 
	for observability recovery. 
\end{cor}
	See the proofs in~\cite{TNSE17,commault2011sensor}.
This corollary implies that when a critical measurement  fails, causing loss of observability, measurement of any other state node in the same contraction recovers the system observability~\cite{SPL17}. Recall that (structural) rank deficiency of the system matrix $A$ defines the cardinality of $\delta \mc{M}$ and $\mc{C}$ in its associated system graph~$\mc{G}_A$~\cite{murota,icassp13}.

\section{Contraction Prevalence in SF Graphs} \label{sec_SF}
In this section, the distribution of contractions in SF networks, as random graph-representations of real-world systems, is studied. Such random models simplify the understanding of different processes, e.g., spreading processes and cascading failures~\cite{newman2006structure,Holme2002clusteringScaleFree,turker2018generating,herrero2015ising}. The main feature of SF graphs is their  \textit{power-law degree distribution}~\cite{newman2006structure}, which implies that there are few hubs (nodes with high degree) and large number of low-degree nodes in the SF network. To construct such networks, Ref.~\cite{newman2006structure} provides an iterative algorithm initializing with a small \textit{seed graph} and recursively adding a new node with $m$ new edges. The main feature of this iterative procedure is that the linking probability between the new node and an old node is proportional to its degree. In other words, the new node \textit{prefers} connecting to old hubs, hence it is named \textit{preferential attachment}. Such SF graphs are known to have low GCC\footnote{GCC is defined as the ratio of the triangles $tr$ to the total number of connected open triplets $trp$ in the graph, i.e., $\text{GCC} = \frac{3.tr}{trp}$~\cite{newman2006structure}.}, while real-world networks, for example social networks, show high clustering. Therefore, a modified model with high GCC is proposed \cite{Holme2002clusteringScaleFree,turker2018generating,herrero2015ising}, named Clustered Scale Free (CSF). The building blocks of this model are similar to the preferential attachment, where the difference is the \textit{triad formation} step. In this model, the newly added node directly links to 
$m_r$ nodes, while also making $m_s$ preferential linking to some neighbors of  $m_r$ preferentially attached nodes to create triads. This significantly increases the GCC in CSF networks with the same average node degree as  SF networks.

\subsection{Empirical results and simulation}
\begin{figure}[hbpt!]
	\centering
	\includegraphics[width=1.65in]{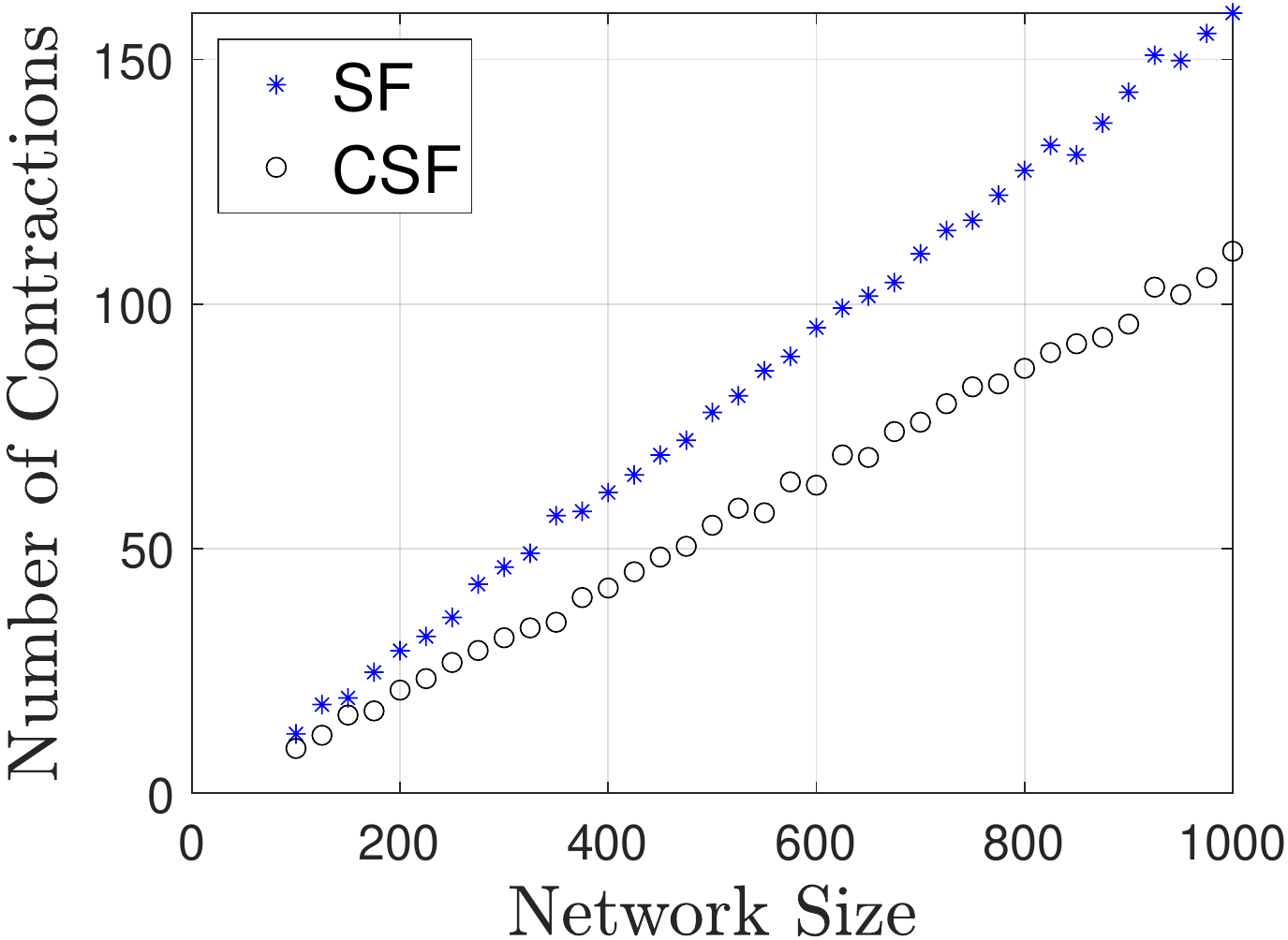}
	\includegraphics[width=1.65in]{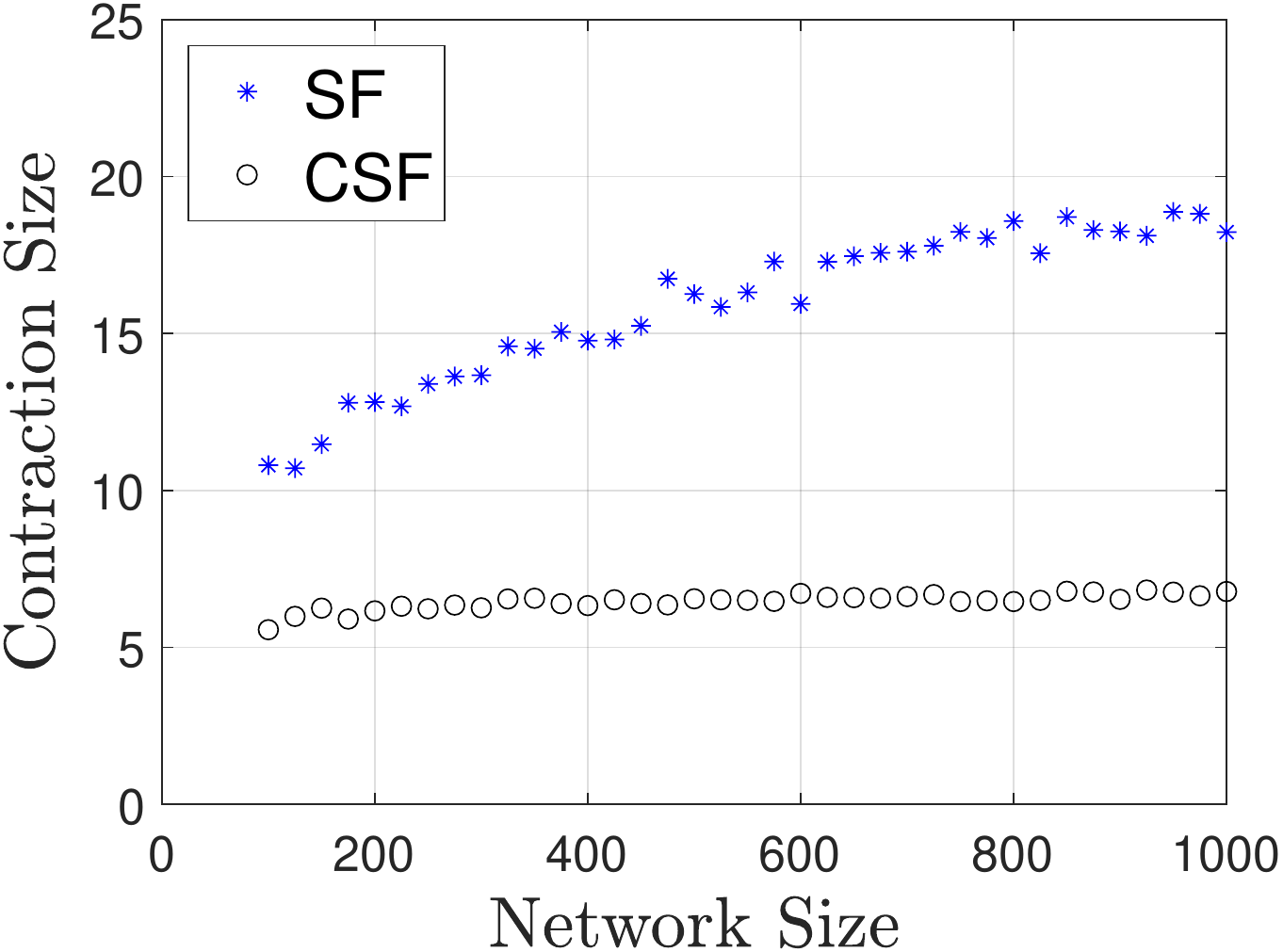}
	\caption{Number and size of contractions averaged over $50$ realizations of growing SF and CSF networks.}
	\label{fig_SF_CSF}
\end{figure}

To study the effect of GCC, we  compare the  number/average-size  of contractions in CSF/SF graphs. We perform Monte-Carlo simulations over $50$ realizations of sample CSF and SF graphs with ${m=m_r+m_s=2}$ and $n=100$ to $n=1000$ nodes. 
Having $m=m_r+m_s$ ensures equal number of new edges via preferential attachment in both CSF and SF networks, implying the same average node degrees for both types. This is essential for comparison  as all features of both CSF and SF graphs must be similar while only their GCC differs~\cite{Holme2002clusteringScaleFree}. Fig.~\ref{fig_SF_CSF} shows the  Monte-Carlo simulation results.

As shown in Fig.~\ref{fig_SF_CSF}, in SF graphs there are more contractions  which are in average larger (in size) as compared to CSF graphs. Table~\ref{tab_SF} summarizes the results shown in Fig.~\ref{fig_SF_CSF}.
\begin{table}[hbpt!]
	\centering
	\caption{Comparison of  CSF and SF graph samples.}
	\begin{tabular}{|l|c|c|}
		\hline
		Graph Type&~ SF &~CSF\\
		\hline
		Average size of Contraction & ~$15.8$  &~$6.6$ \\
		\hline
		Average number of Contractions & ~$86$  &~$59$ \\
		\hline
		Average GCC & ~$0.017$  &~$0.181$ \\
		\hline
		\hline
	\end{tabular}
	\label{tab_SF}
\end{table}


\subsection{Discussions on the results}
From Fig.~\ref{fig_SF_CSF} and Table~\ref{tab_SF},  we see the average size of contractions in SF networks is significantly larger than CSF networks. Recall that both  graph types are constructed via the preferential attachment method and, therefore, both share similarity of most graph properties, e.g., (i) logarithmic growth in mean shortest-path and (ii) power-law degree distribution~\cite{Holme2002clusteringScaleFree}. Their main difference stems from the GCC, which is lower in SF graphs. This is the key feature contributing to the decrease in both size and number of contractions in CSF graphs. Again we emphasize that the other graph properties of both types are similar. Thus, one can conclude that, in graphs with power-law degree distribution, increase in GCC causes fewer contractions with smaller average-size.

In terms of system observability/estimation, the implication is that for graphs with higher GCC: (i) fewer state measurements are needed to ensure observability; and (ii) fewer observationally equivalent states are available to recover observability loss in case of measurement failure. The first result
deduced from number of contractions while the latter stems from average size of contractions. This implies that the observability (and consequently estimation properties) can be improved/deteriorated by tuning the GCC of (synthetic) system networks via~\cite{csl2020,islam2018towards,moore2021inclusivity}. A such example is given next.


\section{Illustrative Example: Application to Power Networks} \label{sec_power}
\begin{figure}[t]	
	\centering
	\includegraphics[width=3in]{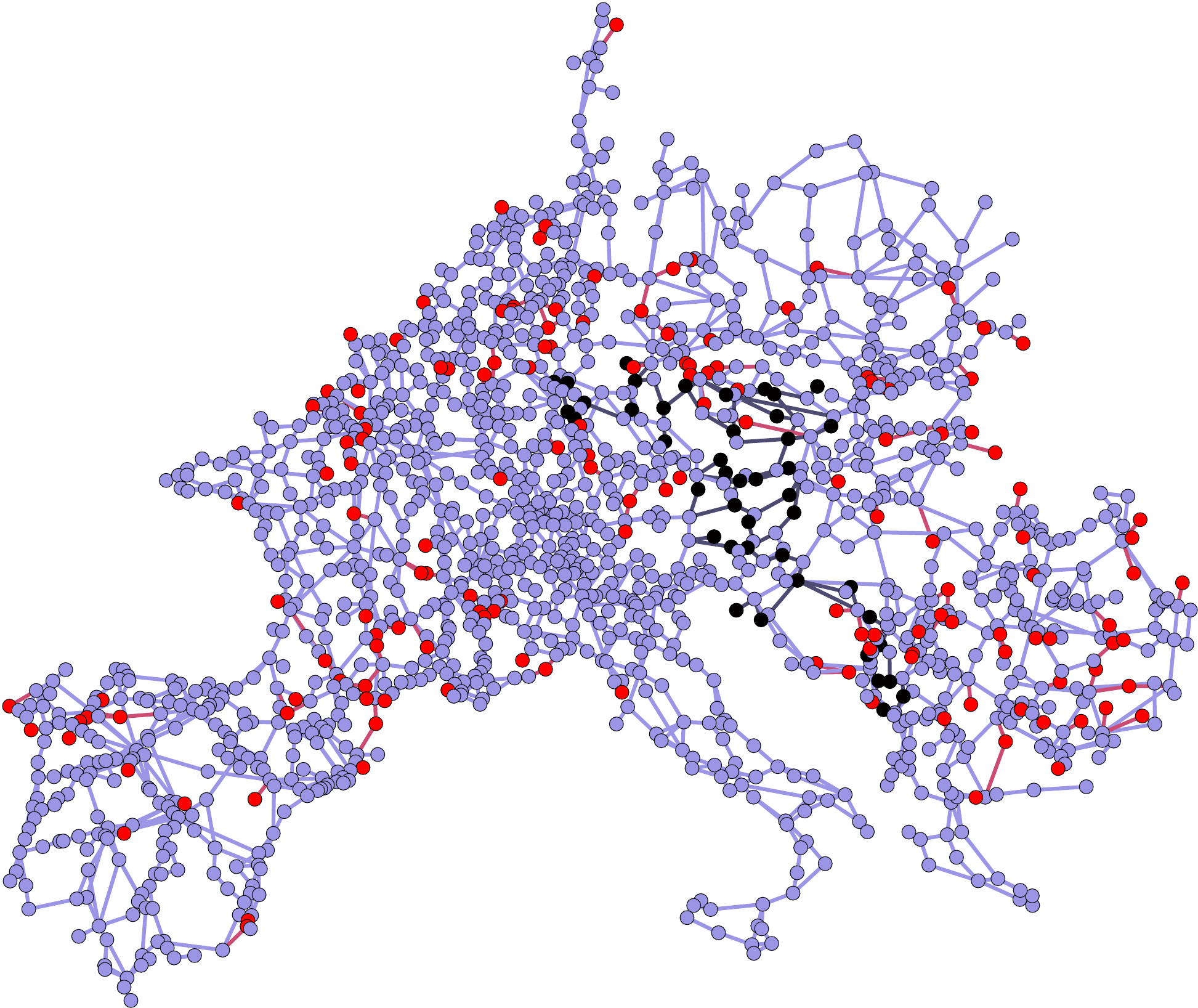}
	\includegraphics[width=3in]{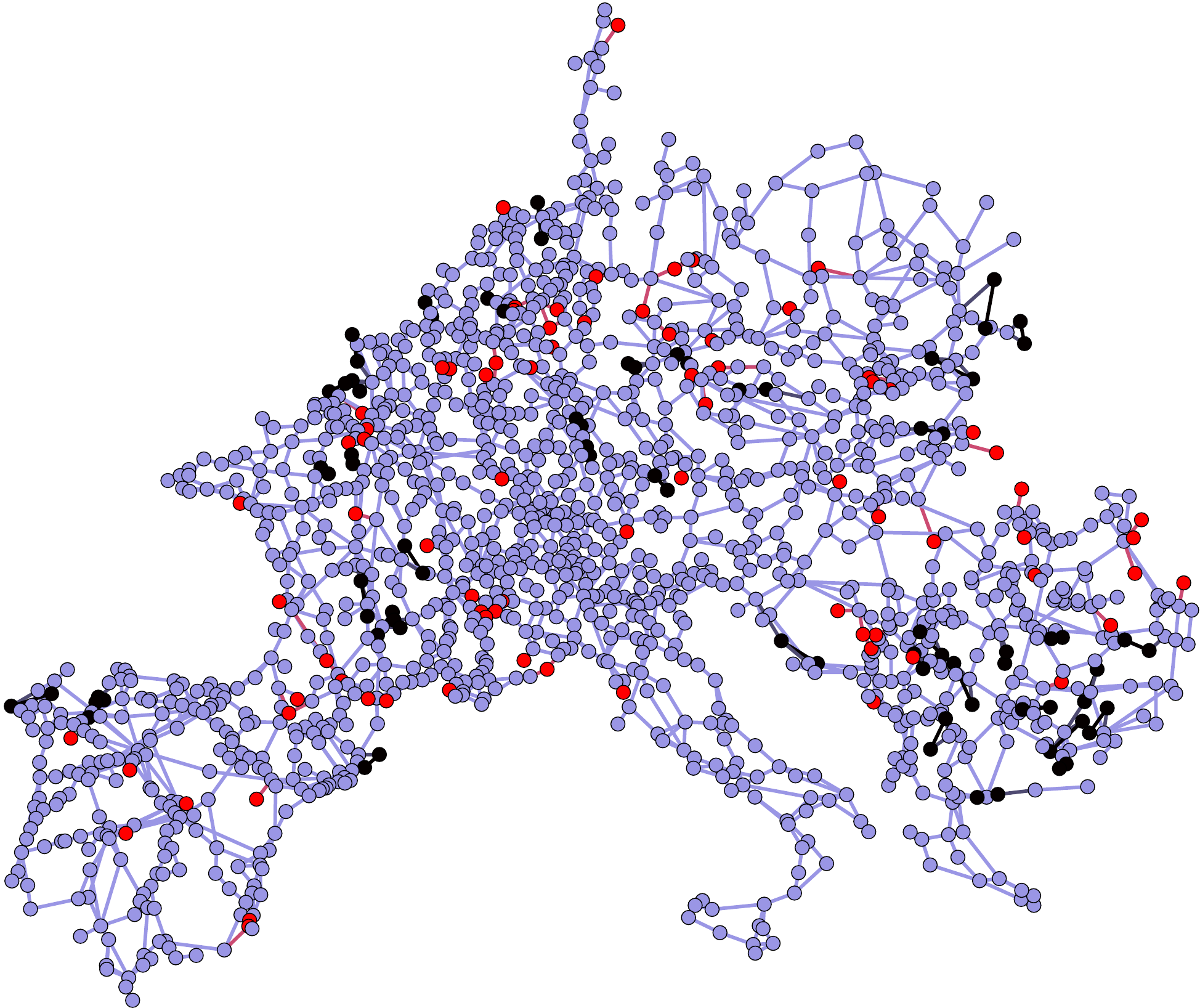}	
	\caption{(top) The European power grid with $151$ unmatched bus-nodes (shown in red) whose states need to be measured via PMUs. An example contraction of $46$ nodes is also shown in black. (bottom) By adding $40$ edges (shown in black) to increase the GCC of power network, the number of unmatched bus-nodes (and minimum required PMUs) is reduced to $80$.  }\label{fig_grid}
\end{figure}

The power grid, as an engineered infrastructural network, can be conceived as a large-scale dynamical system~\cite{machowski2020power}, where the sparcity of its (linearized) system matrix follows the structure of the power distribution network~\cite{camsap11}. To ensure reliable power delivery, the electrical phasor state (voltage, current, etc.) is typically measured via advanced sensors, known as phasor measurement units (PMUs), distributed over the electricity grid. The PMU placement is such that to ensure observability of the entire power grid for monitoring puposes~\cite{babu2016optimal}. Following the results of this paper, one can reduce the number of allocated PMUs for  observability by changing the structure of the power network.    
Consider the European power grid~\cite{power_data}  shown in Fig.~\ref{fig_grid}(top) with $1494$ nodes (buses) and $2156$ edges (transmission lines).  The unmatched nodes represent the possible location of PMU placement in the grid. 
Following the results in Section~\ref{sec_SF}, 
we increase the GCC by adding $40$ edges between certain bus-nodes in the network as shown in Fig.~\ref{fig_grid}(bottom). The network grid properties before and after link addition are compared in Table~\ref{tab_power}. Note that the change in average node degree is negligible while the GCC is increased about $19\%$. As expected, contractions and unmatched nodes are reduced by $47\%$ by adding only $40$ edges which are about $1.8\%$ of the total edges in the  network.  

\begin{table}[hbpt!]
	\centering
	\caption{Network characteristics before/after link addition.}
	\begin{tabular}{|c|c|c|c|}
		\hline
		 links& average degree & GCC & contractions  \\
		\hline
		 $2156$ & $2.886$& $0.094$  &$151$ \\
		\hline
         $2196$& $2.939$ & $0.112$ &$80$  \\
		\hline
		\hline
	\end{tabular}
	\label{tab_power}
\end{table}

\section{Conclusions and Future Research} \label{sec_conc}
In this work, distribution of contractions in SC digraphs and its relation with graph GCC is studied. Note that for general non-SC digraphs, another component known as \textit{root SCC} or \textit{parent SCC} also affects system graph observability \cite{pequito2017trade,globalsip14}.
As future research, we intend to study the correlation between  prevalence/size of both parent SCCs and contractions with other graph features, such as assortativity/disassortativity and community/hierarchical structure~\cite{newman2006structure}.

\bibliographystyle{IEEEbib}
\small{\bibliography{bibliography}}
\end{document}